 \documentclass[smallabstract,smallcaptions]{dccpaper}

\usepackage{graphicx}
\usepackage{citesort}
\usepackage{amsmath}
\usepackage{amssymb}
\usepackage{color}
\usepackage{url}
\usepackage{algorithm}
\usepackage{algpseudocode}
\usepackage{makecell}

\newlength{\figurewidth}
\newlength{\smallfigurewidth}

\newcommand{\BWT}{\ensuremath{\mathrm{BWT}}}
\newcommand{\LF}{\ensuremath{\mathrm{LF}}}
\newcommand{\SA}{\ensuremath{\mathrm{SA}}}
\newcommand{\LCP}{\ensuremath{\mathrm{LCP}}}
\newcommand{\LCE}{\ensuremath{\mathrm{LCE}}}
\newcommand{\MS}{\ensuremath{\mathrm{MS}}}
\newcommand{\pos}{\ensuremath{\mathrm{pos}}}
\newcommand{\len}{\ensuremath{\mathrm{len}}}
\newcommand{\select}{\ensuremath{\mathrm{select}}}
\newcommand{\rank}{\ensuremath{\mathrm{rank}}}

\begin{document}

\title
{\large
\textbf{Augmented Thresholds for MONI}
}

\author{%
C\'esar Mart\'inez-Guardiola$^{\ast}$, Nathaniel K. Brown$^{\dag}$, Fernando Silva-Coira$^{\ast}$,\\Dominik K\"{o}ppl$^{\ddag}$, Travis Gagie$^{\dag}$ and Susana Ladra$^{\ast}$ \\[0.5em]
{\small\begin{minipage}{\linewidth}\begin{center}
\begin{tabular}{ccccc}
$^{\ast}$Universidade da Coru\~na & \hspace*{0.25in} & $^{\dag}$Dalhousie U & \hspace*{0.25in} & $^{\ddag}$TMDU \\
CITIC, A Coru\~na, Spain && Halifax, Canada && Tokyo, Japan \\
\url{{first.last}@udc.es} && \url{{first.last}@dal.ca} && \url{koeppl.dsc@tmd.ac.jp} \\
\end{tabular}
\end{center}\end{minipage}}}

\maketitle
\thispagestyle{empty}

\begin{abstract}
MONI (Rossi et al., 2022) can store a pangenomic dataset $T$ in small space and later, given a pattern $P$, quickly find the maximal exact matches (MEMs) of $P$ with respect to $T$.  In this paper we consider its one-pass version (Boucher et al., 2021), whose query times are dominated in our experiments by longest common extension (LCE) queries.  We show how a small modification lets us avoid most of these queries and thus significantly speeds up MONI in practice while only slightly increasing its size.
\end{abstract}

\section{Introduction}
\label{sec:introduction}

The FM-index~\cite{FM00} is one of the most successful compact data structures and DNA alignment has been its ``killer app'', with FM-based aligners such as Bowtie~\cite{LTPS09,LS12} and BWA~\cite{LD09} racking up tens of thousands of citations and seeing every day use in labs and clinics worldwide.  Standard FM-indexes can handle only a few human genomes at once, however, and geneticists now realize that aligning against a only few standard references biases their research results and medical diagnoses~\cite{Beg19}.  Among other concerns, this bias undermines personalized medicine particularly for people from ethnic groups --- such as African, Central/South Asian, Indigenous, Latin American and Middle Eastern populations --- whose genotypes are not reflected well in the standard references or even in public databases of genomes~\cite{LAWRB18}.  Countries such as China~\cite{WWLLTGFZLZ+08} and Denmark~\cite{MJPSLVSBTI+17} have assembled their own reference sequences, but it is not clear whether and how we can do this fairly for multi-ethnic populations.  Bioinformaticians and data-structure designers have therefore been looking for ways to index models that better capture the genetic diversity of whole species, especially humanity.  The most publicized approach so far is building and indexing pangenome graphs~\cite{SMCNEMSHCC+21}, but we can also try scaling FM-indexes up to handle a dozen or so representative samples genomes~\cite{CSMIL21} or, more ambitiously, to handle even thousands of genomes at once.  Indexing thousands of genomes at once is technically challenging, of course, but it should give us different functionality than pangenome graphs.

M\"akinen et al.~\cite{SVMN08,MNSV10} initiated the study of indexing massive genomic datasets with their index based on the run-length compressed Burrows-Wheeler Transform (RLBWT), which stores such a pangenomic dataset $T [1..n]$ in space proportional to the number $r$ of runs in the BWT of $T$ and allows us to quickly {\em count} the number of exact matches of any pattern $P [1..m]$ in $T$.  Policriti and Prezza~\cite{PP18} showed that, if we augment M\"akinen et al.'s index with the entries of the suffix array (SA) sampled at BWT run boundaries, then we can quickly locate {\em one} of $P$'s matches in $T$.  Gagie, Navarro and Prezza~\cite{GNP20} then showed how we can store that SA sample such that we can quickly locate {\em all} of $P$'s matches in $T$.  (For the sake of brevity, we assume the reader is familiar with the BWT, SA, etc.; otherwise, we refer them to M\"akinen et al.'s~\cite{MBCT15} and Navarro's~\cite{Nav16} texts.)  Gagie et al.\ called their data structure the $r$-index, after its $O (r)$ space bound; Nishimoto and Tabei~\cite{NT21} recently sped it up to answer queries in optimal time when $T$ is over a $\mathrm{polylog} (n)$-sized alphabet, while still using $O (r)$ space.  Boucher et al.~\cite{BGKLMM19,KMBGLM20} showed how we can build an $r$-index efficiently in practice using a technique they called prefix-free parsing (PFP).

Because approximate pattern matching is often more important in bioinformatics than exact matching, Bannai, Gagie and I~\cite{BGI20} designed a version of the $r$-index that can efficiently find maximal exact matches (MEMs), which are commonly used for approximate pattern matching in tools such as BWA-MEM~\cite{Li13}.  Bannai et al.'s is not a true $r$-index because it requires fast random access to $T$ and we do not know how to support that in worst-case $O (r)$ space, but Gagie et al.~\cite{GMNST19,GMNSST20} showed how we can use PFP to build a straight-line program (SLP) for $T$ that gives us this random access and in practice takes significantly less space than the $r$-index itself.  The key idea behind Bannai et al.'s index is to store the positions of $r$ thresholds in the RLBWT, one between each consecutive pair of runs of the same character, but they did not give an algorithm for finding those thresholds.  Rossi et al.~\cite{ROLGB22} showed that we can choose the thresholds based on the longest common prefix (LCP) array and build Bannai et al.'s index efficiently with PFP.  They implemented it in a tool called MONI (Finnish for ``multi'', as it indexes many genomes at once), and demonstrated its practicality for pangenomic alignment.

By default, MONI makes two passes over $P$, one right-to-left and then the other left-to-right.  Boucher et al.~\cite{BGIKLMNPR21} noted, however, that by using the SLP to support longest common extension (LCE) queries instead of random access, MONI can run in one pass.  For long patterns, MONI in two-pass mode buffers a significant amount of data during its first pass, so switching to one-pass mode reduces its workspace and allows us to run more queries in parallel.  We can also use one-pass MONI for applications that are inherently online, such as recognizing and ejecting non-target DNA strands from nanopore sequencers~\cite{ARKSGBL21}. Even though one-pass MONI processes most characters in $P$ without LCE queries, the LCE queries it does compute still take most of the query time~\cite{BGIKLMNPR21}.  We show in this paper that by precomputing and storing two LCE values for each threshold, in practice we can avoid many of those queries and thus significantly speed up one-pass MONI while increasing its size only slightly.

\section{MONI}
\label{sec:MONI}

Bannai et al.\ defined a {\em threshold} between two consecutive runs $\BWT [s_1..e_1]$ and $\BWT [s_2..e_2]$ of the same character, to be a position $t$ with $e_1 < t \leq s_2$ such that $\LCE (e_1, k) \geq \LCE (k, s_2)$ for $k < t$, and $\LCE (e_1, k) \leq \LCE (k, s_2)$ for $k \geq t$.  (Rossi et al.'s construction is based on the observation that we can set $t$ to the position of a minimum in $\LCP [e_1 + 1..s_2]$.)  Bannai et al.\ showed how adding these thresholds to an $r$-index lets us compute MEMs by computing the matching statistics $\MS [1..m]$ of $P$ with respect to $T$, where the $i$th matching statistics $\MS [i].\pos$ and $\MS [i].\len$ are defined such that
\[T \left[ \rule{0ex}{2ex} \MS [i].\pos..\MS [i].\pos + \MS [i].\len - 1 \right]
= P [i..i + \MS [i].\len - 1]\]
and $P [i..i + \MS [i].\len]$ does not occur in $T$.  In other words, $\MS [i].\pos$ is a pointer to the starting position in $T$ of a longest match for $P [i..m]$ and $\MS [i].\len$ is the length of that match, where a {\em longest match} for $P [i..m]$ is an occurrence in $T$ of the longest prefix of $P [i..m]$ that occurs in $T$.

Suppose we have already computed $\MS [i + 1].\pos$ and the position $j$ of $T [\MS [i + 1].\pos - 1]$ in the BWT.  If $\BWT [j] = P [i]$, then $\MS [i].\pos = \MS [i + 1].\pos - 1$ and we can continue once we compute the position $\LF (j)$ of $T [\MS [i].\pos - 1]$ in the BWT.  Otherwise, let $\BWT [e]$ be the last occurrence of $P [i]$ before $\BWT [j]$, and $\BWT [s]$ be the first occurrence of $P [i]$ after $\BWT [j]$.  By the definitions of the BWT and thresholds, if $\BWT [j]$ is strictly above the threshold between $\BWT [e]$ and $\BWT [s]$, then a prefix of $T [\SA [e]..n]$ is a longest match for $P [i..m]$; otherwise, a prefix of $T [\SA [s]..n]$ is a longest match for $P [i..m]$.  Since $\BWT [e]$ is the end of a run and $\BWT [s]$ is the start of a run, we have $\SA [e]$ and $\SA [s]$ stored.  Therefore, depending on whether $\BWT [j]$ is above or below the threshold, either we can ``jump up'' from $\BWT [j]$ to $\BWT [e]$ and set $\MS [i].\pos = \SA [e]$ (so the position of $T [\MS [i].\pos - 1]$ in the BWT is $\LF (e)$), or we can ``jump down'' from $\BWT [j]$ to $\BWT [s]$ and set $\MS [i].\pos = \SA [e]$ (so the position of $T [\MS [i].\pos - 1]$ in the BWT is $\LF (s)$).

By default, MONI makes a right-to-left pass over $P$ to compute $\MS [1..m].\pos$, and then a left-to-right pass over $P$ to compute $\MS [1..m].\len$.  If we use the SLP to support LCE queries instead of random access, however, then we need only one pass over $P$.  To see why, suppose that when we compute $\MS [i].\pos$, we have already computed $\MS [i + 1].\len$ as well as $\MS [i + 1].\pos$.  If we jump up from $\BWT [j]$ to $\BWT [e]$, then
\begin{eqnarray}
\MS [i].\len
& = & \min \left( \rule{0ex}{2ex} \LCE (\MS [i + 1].\pos, \SA [e]), \MS [i + 1].\len \right) + 1\,;
\label{eqn:jump_up}
\end{eqnarray}
if we jump down from $\BWT [j]$ to $\BWT [s]$, then
\begin{eqnarray}
\MS [i].\len
& = & \min \left( \rule{0ex}{2ex} \LCE (\MS [i + 1].\pos, \SA [s]), \MS [i + 1].\len \right) + 1\,.
\label{eqn:jump_down}
\end{eqnarray}
In fact, if we compute both $\LCE (\MS [i + 1].\pos, \SA [e])$ and $\LCE (\MS [i + 1].\pos, \SA [s])$, then we need not check the threshold between $\BWT [e]$ and $\BWT [s]$ at all.  MONI stores the thresholds in order to use only one LCE query for each jump, because the thresholds collectively do not take much space compared to the RLBWT and the SA samples, and the LCE queries are slow compared to the LF-steps.

\section{Augmented Thresholds}
\label{sec:augmented_thresholds}

In practice, MONI's jumps and resultant LCE queries tend to occur in bunches: if a character $P [i]$ is a sequencing error or a variation not in $T$, then we will probably jump for $P [i]$, find a short longest match, and then also jump for several more characters of $P$ in rapid succession, until the longest matches are finally long enough again to reorient us in the BWT.  Because the lengths of the longest matches can only increment for each character of $P$ we process, most of the comparisons in Equations~\ref{eqn:jump_up} and~\ref{eqn:jump_down} between the LCE values and the length of the current longest match will simply return the length of the current match.  This observation led us to wonder if all those LCE queries are really necessary.

\begin{figure}[ht]
\begin{center}
\begin{tabular}{r@{\hspace{3ex}}c@{\hspace{3ex}}c@{\hspace{3ex}}l}
   $k$\ \ \  & $\SA [k]$ & $\BWT [k]$ & $T [\SA [k]..n]$\\
\hline
\multicolumn{1}{c}{\ \ \ \ $\vdots$}  &
\multicolumn{1}{c}{$\vdots$\ \ \ \ }  &
\multicolumn{1}{c}{$\vdots$\ \ \ \ }  &
\multicolumn{1}{c}{$\vdots$\ \ \ }\\
       1234  & 8765      & \tt A      & \tt GAGACATCA\dots\\
$e_1 = 1235$ & 1519      & \tt A      & \tt \textcolor{red}{GAT}\textcolor{green}{A}CATTA\dots\\
       1236  & 5450      & \tt C      & \tt \textcolor{red}{GAT}\textcolor{green}{A}GATTA\dots\\
  $j = 1237$ & 1004      & \tt G      & \tt \textcolor{red}{GAT}\textcolor{green}{A}TAGAA\dots\\
       1238  & 4242      & \tt G      & \tt \textcolor{red}{GAT}CCAATA\dots\\
  $t = 1239$ & 3110      & \tt G      & \tt \textcolor{blue}{GATTA}CATA\dots\\
       1240  & 1102      & \tt T      & \tt \textcolor{blue}{GATTA}CTTA\dots\\
       1241  & 1978      & \tt T      & \tt \textcolor{blue}{GATTA}GATA\dots\\
$s_2 = 1242$ & 2505      & \tt A      & \tt \textcolor{blue}{GATTA}TCAT\dots\\
       1243  & 2022      & \tt A      & \tt GATTATGAA\dots\\
\multicolumn{1}{c}{\ \ \ \ $\vdots$}  &
\multicolumn{1}{c}{$\vdots$\ \ \ \ }  &
\multicolumn{1}{c}{$\vdots$\ \ \ \ }  &
\multicolumn{1}{c}{$\vdots$\ \ \ }
\end{tabular}
\end{center}
\caption{Suppose we want to compute $\MS [i].\len$ for some $i$ such that $P [i] = \BWT [e_1] = \BWT [s_2]$, and the position $j$ of $T [\MS [i + 1].\pos - 1]$ in the BWT is between $e_1 + 1$ and $t - 1$.  If $\MS [i + 1].\len \leq \LCE (\SA [e_1], \SA [t - 1])$ then, since $\LCE (\SA [e_1], \SA [t - 1]) \leq \LCE (\SA [e_1], \SA [j])$, by transitivity $\MS [i + 1].\len \leq \LCE (\SA [e_1], \SA [j])$ and we can safely set $\MS [i].\len = \MS [i + 1].\len + 1$.}
\label{fig:example}
\end{figure}

Suppose that, at the threshold $t$ between between two consecutive runs $\BWT [s_1..e_1]$ and $\BWT [s_2..e_2]$ of the same character, we store $\LCE (\SA [e_1], \SA [t - 1])$ and $\LCE (\SA [t], \SA [s_2])$.  Furthermore, suppose we later want to compute $\MS [i].\len$ for some $i$ such that $P [i] = \BWT [e_1] = \BWT [s_2]$ and the position $j$ of $T [\MS [i + 1].\pos - 1]$ in the BWT is between $e_1 + 1$ and $s_2 - 1$.  If $j < t$ and
\[\MS [i + 1].\len \leq \LCE (\SA [e_1], \SA [t - 1])\,,\]
or $j \geq t$ and
\[\MS [i + 1].\len \leq \LCE (\SA [t], \SA [s_2])\]
then, as illustrated in Figure~\ref{fig:example}, we can safely set $\MS [i].\len = \MS [i + 1].\len + 1$ without using an LCE query. Algorithm~\ref{alg:augmented} shows how these values are used to compute $\MS [1..m]$ for a given pattern $P[1..m]$ by storing the thresholds alongside these ``threshold LCEs''.

\begin{algorithm}[!ht]
\caption{Computes \MS\ using a variation of one-pass MONI~\cite{BGIKLMNPR21} which stores augmented thresholds (thresholds and thr\_lce arrays)}
\label{alg:augmented}
\begin{algorithmic}[1]
\State $j \gets \BWT.\select_{P[m]}(1)$
\State $\MS[m] \gets (\pos: \SA[j], \len: 1)$
\For {$i = m-1$ \textbf{down to} $1$}
    \If {$\BWT[j] = P[i]$}
        \State $\MS[i] \gets (\pos: \MS[i+1].\pos - 1, \len: \MS[i+1].\len + 1)$
    \Else
        \State $c \gets \BWT.\rank_{P[i]}(j)$
        \State $e_1 \gets \BWT.\select_{P[i]}(c)$ 
        \State $s_2 \gets \BWT.\select_{P[i]}(c + 1)$ 
        \State $x \gets \BWT.\text{run\_of\_position}(s_2)$ \Comment{Position $s_2$ belongs to the $x$th run}
        \State $t \gets \text{thresholds}[x]$
        \If {$j < t$}
            \Comment $\text{thr\_lce}_e$ stores $\LCE(\SA[e_1], \SA[t - 1])$
            \If {$\MS[i+1].\len \leq \text{thr\_lce}_e[x]$}
                \State $\MS[i].\len \gets \MS[i+1].\len + 1$
            \Else
                \State $\MS[i].\len \gets min(\MS[i+1].\len, \LCE(\SA[e_1], \MS[i+1].\pos) + 1$
            \EndIf
            \State $\MS[i].\pos \gets \SA[e_1]$
            \State $j \gets \LF(e_1)$
        \Else 
            \Comment $\text{thr\_lce}_s$ stores $\LCE(\SA[t], \SA[s_2])$
            \If {$\MS[i+1].\len \leq \text{thr\_lce}_s[x]$}
                \State $\MS[i].\len \gets \MS[i+1].\len + 1$
            \Else
                \State $\MS[i].\len \gets min(\MS[i+1].\len, \LCE(\SA[s_2], \MS[i+1].\pos) + 1$
            \EndIf
            \State $\MS[i].\pos \gets \SA[s_2]$
            \State $j \gets \LF(s_2)$
        \EndIf
    \EndIf
\EndFor
\end{algorithmic}
\end{algorithm}

Threshold LCEs can be computed using \LCE\ queries and SA samples, but their relationship to thresholds allows us to compute both simultaneously. Recall that Rossi et al. observed that we can set $t$ to be the position of $min(\LCP[e_1+1..s_2])$ using a range-minimum query (RMQ), which they support space-efficiently through PFP and a range-minimum data structure over the \LCP\ array~\cite{ROLGB22}. We can also define \LCE\ queries as RMQs over the \LCP\ array~\cite{INT10}, such that $\LCE(\SA[e_1], \SA[t - 1]) = min(\LCP[e_1..t-1])$ and $\LCE(\SA[t], \SA[s_2]) = min(\LCP[t+1..s_2])$. These minimums can be computed alongside the thresholds by performing RMQs for the given ranges as thresholds are found. This operation scans each run boundary and with only a slight modification to the original MONI method builds both the thresholds and the threshold LCEs (constituting augmented thresholds).

\section{Experiments}
\label{sec:experiments}

We directly compare the time and memory for querying the augmented thresholds approach against the unmodified one-pass MONI. To mitigate the size increase of augmented thresholds, we explore techniques for space-efficiency. Any single threshold LCE can be stored in $O(\lg{n})$-bits (since they inherit \LCP\ bounds); however, many values tend to be smaller than others~\cite{KKP16} and in practice our LCE values represent minimums over ranges of the \LCP\ array. The second observation is the existence of threshold LCEs which can be ignored: if $t = s_2$ then for any position $j$ (with $e_1 < j < s_2$) we always have $j < t$ so we jump up to $e_1$ and the corresponding \LCE\ is never used, and similarly for $t = e_1+1$ and always jumping down. Thus, we can safely ignore these values, choosing to ``zero'' them or not store any value at all. For thresholds, we note that they form increasing sub-sequences with respect to each of the $\sigma$ unique characters in the text; we compress the thresholds by storing them in $\sigma$ bitvectors as done in Ahmed et al.'s implementation~\cite{ARKSGBL21}. 

We focus on selected variants of augmented thresholds which differ in storing the threshold LCEs and compare against the unmodified approach:
\begin{itemize}
    \setlength{\itemsep}{0pt}
    \setlength{\parskip}{0pt}
    \setlength{\parsep}{0pt}
    \item \texttt{PHONI}: Standard version of one-pass MONI described as \texttt{PHONI}$_{std}$ in original paper~\cite{BGIKLMNPR21}.
    \item \texttt{Aug-Full}: One-pass MONI modified with augmented thresholds described previously, using $O(\lg{n})$-bits per threshold LCE stored.
    \item \texttt{Aug-1}: As above, but caps the size to one byte per threshold LCE. In the event of an overflow, we default to performing a single \LCE\ query.
    \item \texttt{Aug-BV-Full}:  Stores a bitvector marking which threshold LCEs are used/non-zero, storing just these values with $O(\lg{n})$-bits for each.
    \item \texttt{Aug-BV-1}: As above, but ignores storing values greater than one byte (default to \LCE\ query).
    \item \texttt{Aug-DAC}: Stores threshold LCEs using a directly addressable code (DAC) with escaping, as described and tested on the \LCP\ array by Brisaboa et al.~\cite{BLN13}.
    \item \texttt{Aug-BV-DAC}: Same as \texttt{Aug-BV-Full}, but substituting in a DAC to store defined values.
\end{itemize}
Our C++ code is available at \url{https://github.com/drnatebrown/aug_phoni} and is based on the original one-pass MONI code at \url{https://github.com/koeppl/phoni}. All experiments were executed single threaded on a server with an Intel(R) Xeon(R) Bronze 3204 CPU and 512 GiB RAM.

\begin{figure}[!ht]
\begin{minipage}{1\linewidth}
\includegraphics[width=\textwidth]{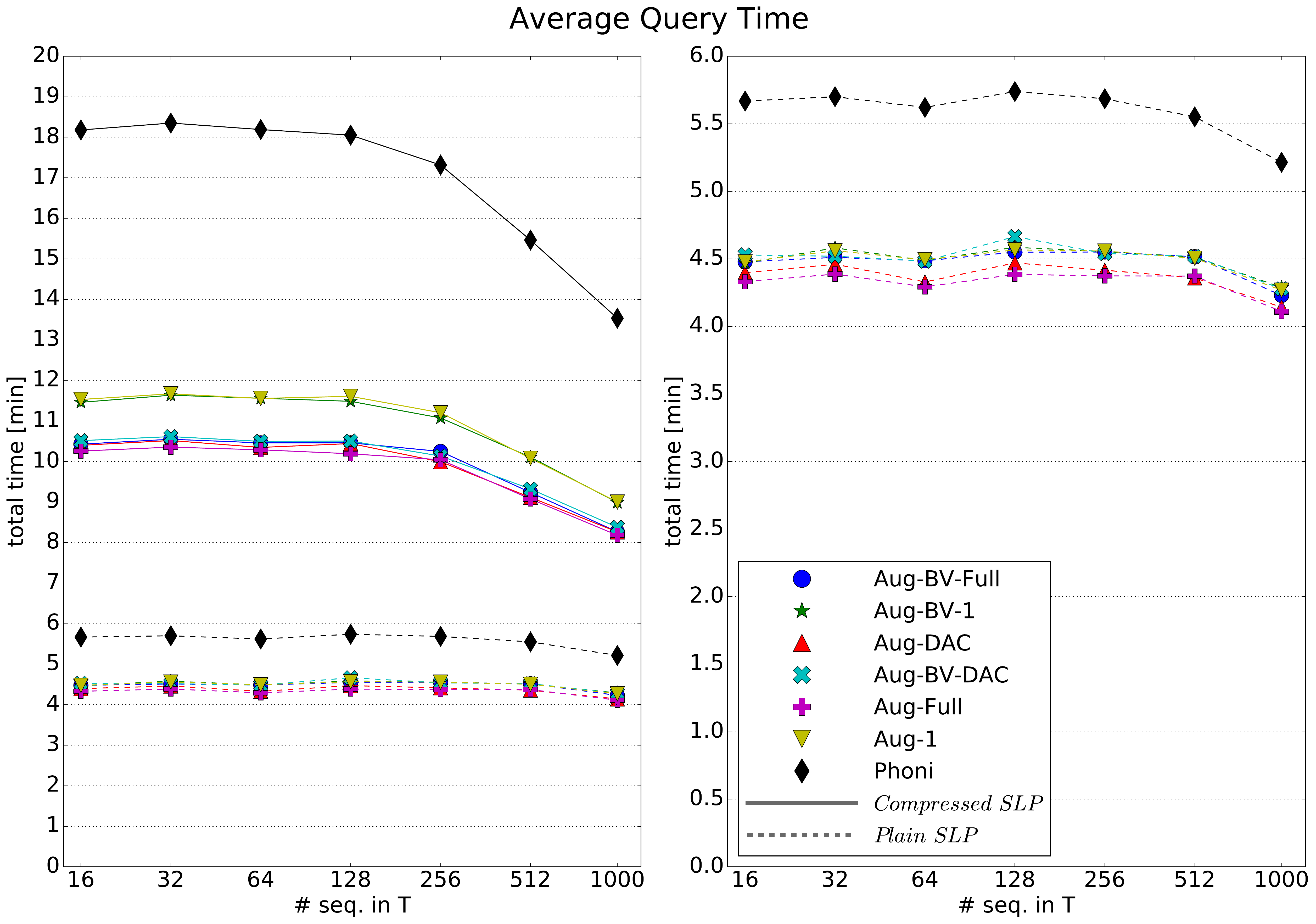}
\end{minipage}
	\caption{The average query time to compute MS using 10 distinct \texttt{chr19} sequences as patterns, using 16, 32, 64, 128, 256, 512, and 1000 sequences of \texttt{chr19} as the text $T$. Data structures shown are as described above. Solid lines use SLP$_{comp}$, dashed lines using SLP$_{plain}$ (focus of right plot).
	\label{fig:query_experiments}}
\end{figure}

\begin{figure}[!ht]
\begin{minipage}{1\linewidth}
\includegraphics[width=\textwidth]{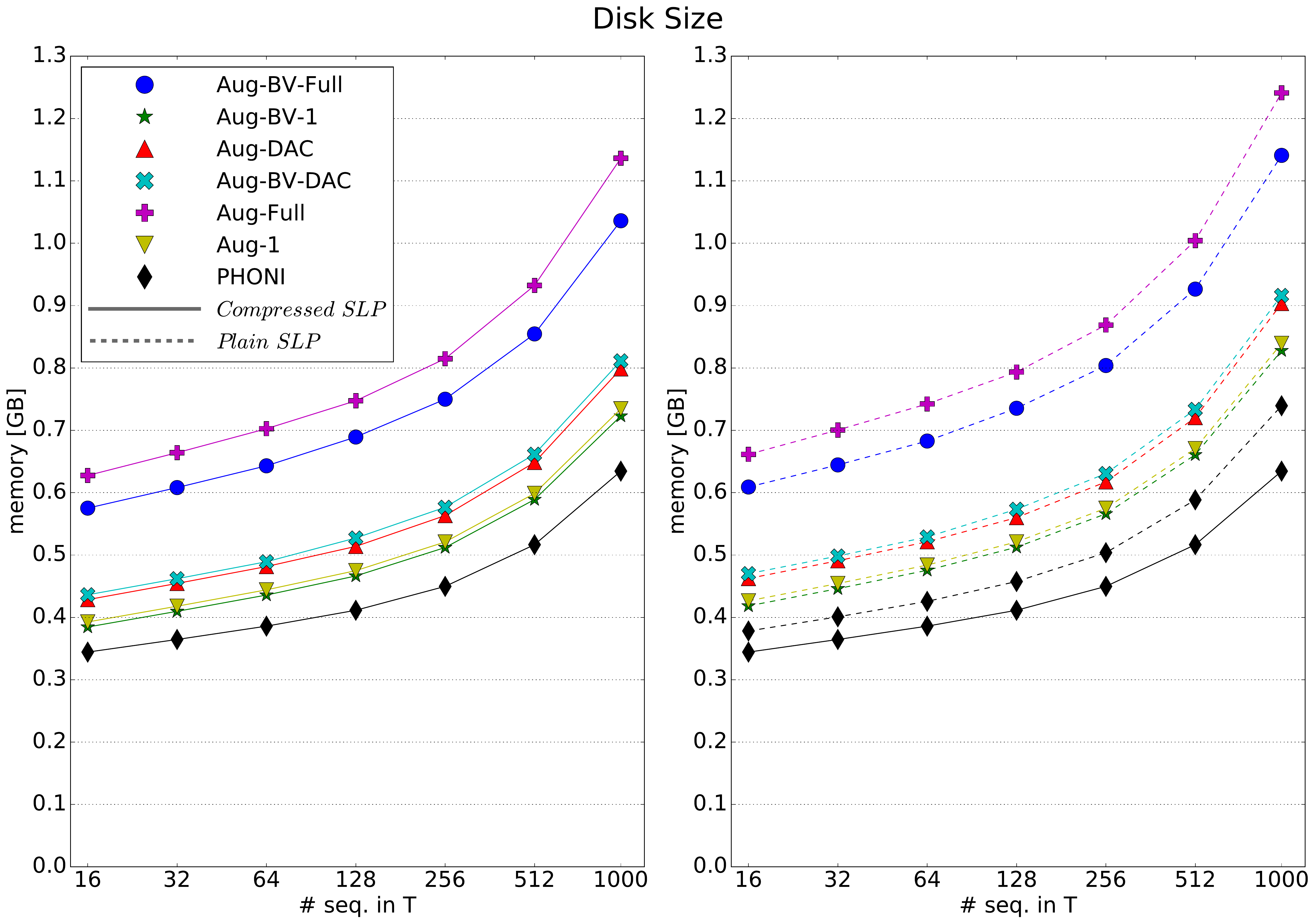}
\end{minipage}
	\caption{The disk size in GB for each data structure built on 16, 32, 64, 128, 256, 512, and 1000 sequences of \texttt{chr19}. Solid lines use SLP$_{comp}$ (focus of left plot), dashed lines using SLP$_{plain}$ (focus of right plot). \texttt{PHONI} using SLP$_{comp}$ is included on right to visualize the size difference of SLP choices.
	\label{fig:size_experiments}}
\end{figure}

To compare against \texttt{PHONI} and its existing results, we re-ran Boucher et al.'s query experiments using the same dataset consisting of chromosome 19 haplotypes (\texttt{chr19}), building the data structures for concatenations of 16, 32, 64, 128, 256, 512, and 1000 sequences of \texttt{chr19} and querying them with 10 different \texttt{chr19} sequences. To support random access and \LCE\ queries efficiently we construct SLPs; both the SLP compressed text of the original one-pass MONI experiments (SLP$_{comp}$), and the naive uncompressed version of Gagie et al. (SLP$_{plain}$)~\cite{GMNSST20} that sacrifices space for speed. The datasets and SLP sizes are reported in Table~\ref{tab:dataset}. The average query times (computing MS for a single pattern) is shown in Figure~\ref{fig:query_experiments} where results for both SLP types are accentuated. Similarly, Figure~\ref{fig:size_experiments} shows the disk sizes for all variants and both SLP types.

\begin{table}[!ht]%
	 		\centering
		 		\begin{tabular}{lrrrrr}
		 			\hline
		 			\# & {$n/10^6$} & {$r/10^4$} & {$n/r$} & ~~SLP$_{comp}$ [MB] & ~~SLP$_{plain}$ [MB] \\ 
		 			\hline
		 			16  &  ~946.01 & 3240.02 & 29.20 & 36.10 & 70.54 \\
		 			32  &  ~1892.01 & 3282.51 & 57.64 & 37.80 & 74.75 \\
		 			64  &  ~3784.01 & 3334.06 & 113.50 & 39.48 & 79.84 \\
		 			128 &  ~7568.01 & 3405.40 & 222.24 & 42.11 & 88.89 \\ 
 		 			256 &  ~15136.04 & 3561.98 & 424.93 & 47.43 & 102.52 \\ 
  		 			512 &  ~30272.08 & 3923.60 & 771.54 & 58.00 & 131.09 \\ 
   		 			1,000 &  ~59125.12 & 4592.68 & 1287.38 & 80.63  & 186.98 \\ 
		 			\hline
		 		\end{tabular}
		 		\bigskip
	 		\caption{Table summarizing the datasets and sizes of SLPs built over them. The first column describes the number of concatenated sequences of \texttt{chr19} representing the text $T$, where $n$ represents the length of $T$ and $r$ the number of runs.\label{tab:dataset}}
	 	\end{table}

\section{Conclusion}
\label{sec:conclusion}

With respect to query times, we can see that any variants using augmented thresholds are always faster than \texttt{PHONI}, and with respect to size, always larger. Introducing the SLP$_{plain}$ clearly benefits all methods by speeding up LCE queries, and although it can be over twice as large as SLP$_{comp}$ when compared directly against each other (Table~\ref{tab:dataset}), the difference is much smaller when comparing the total sizes of the data structures shown in Figure~\ref{fig:size_experiments}. This \LCE\ speedup reduces the gap between query times compared to \texttt{PHONI}, since it spends a larger percentage of execution on them; however, the \LCE\ queries skipped by augmented thresholds still result in faster execution.

We highlight some standout variants when compared to \texttt{PHONI} for the largest text size (1000 sequences of \texttt{chr19}). \texttt{Aug-DAC} is in the fastest class for both SLPs: $48.37\%$ faster and $22.89\%$ larger for SLP$_{comp}$, and $22.92\%$ faster and $19.97\%$ larger for SLP$_{plain}$; significant improvement compared to the original \texttt{PHONI} method (SLP$_{comp}$) and a direct time/space tradeoff for the introduced SLP$_{plain}$. \texttt{Aug-1} is in the smallest class: $40.22\%$ faster and only $14.60\%$ larger for SLP$_{comp}$, while $19.95\%$ faster and $12.66\%$ larger for SLP$_{plain}$. Although \texttt{Aug-Full} is in the fastest class with \texttt{Aug-DAC}, it is much larger. Other variants fall between these approaches in both time and space.

When compared to the original one-pass MONI of Boucher et.\ al (\texttt{PHONI} with SLP$_{comp}$), our best augmented threshold approaches showed over $40\%$ speed improvements with under $20\%$ space increase on the largest dataset, and similar results across all data. When compared to uncompressed threshold LCEs, our applied compression schemes are space-efficient whilst still being faster than unmodified one-pass MONI. Introducing an uncompressed SLP (SLP$_{plain}$) experimentally was shown to be of great benefit to both \LCE\ and total query speed, only requiring a small size increase for computing matching statistics on repetitive texts. Using this SLP, results show augmented thresholds to allow a direct time/space tradeoff (increase speed/space by $\approx20\%$), or a size decrease whilst maintaining a comparable speed increase.

\subsection*{Acknowledgments}


Many thanks to Massimiliano Rossi for guidance on implementing this modification.

\Section{References}
\bibliographystyle{IEEEbib}
\bibliography{refs}

\end{document}